# Long Slot Ridged SIW Leaky Wave Antenna Design Using Transverse Equivalent Technique

Alireza Mallahzadeh, *Member, IEEE*, and Sajad Mohammad-Ali-Nezhad

*Abstract*—A novel ridged SIW leaky-wave antenna (LWA) with controllable side lobe level and low cross-polarization is proposed and fabricated. Longitudinally continuous asymmetric ridges beside SIW sidewalls provide an asymmetric field distribution around the long slot centered on the upper plane of the structure, in way that the slot can radiate. An accurate transverse equivalent network analysis is presented for calculating the properties of the proposed LWA. The wavenumbers of the leaky-mode are calculated theoretically and numerically. The designed ridged SIW LWA shows a good SLL and cross polarization. Measurement results are also consistent with the simulation results.

## I. INTRODUCTION

LEAKY WAVE ANTENNAS (LWAs) exhibit many interesting features in microwave and millimeter-wave applications. Advanced properties of LWAs are frequency scanning capability, wide bandwidth and high directivity with a simple feed network [1]–[3].

Various studies on LWAs have been reported. These investigations include the groove guide [4], [5], the stub loaded rectangular waveguide (LWAs loaded with parallel plates) [6], the nonradiative dielectric guide (NRD) technology [7]–[9], microstrip LWAs [10], [11], half-mode microstrip LWAs [12], a periodic version of the half-mode microstrip LWAs [13] and the strip-loaded dielectric slab [14], [15]. Other LWAs, such as the leaky-wave long slot antenna have also been proposed [16]–[19]. Longitudinal straight leaky-wave long slot antennas provide narrow elevation beamwidth, low profile, and high power capacity.

The conventional straight slot antenna in which the aperture field distribution is uniform has a poor side lobe level (SLL). The desired far-field pattern can be achieved by changing the width of the slot or its distance from the center. While properly tapering the aperture distribution improves SLL, it can cause an increase in cross-polarization [17]. Cross-polarization increases in LWAs that lack a symmetrical aperture distribution, such as those with curved slots. In LWA structures, parallel plate stubs are used for decreasing the cross-polarization. Different mechanisms for cross-polarization reduction are discussed in [20].

In microwave and millimeter-wave applications, waveguide devices are preferred to microstrip devices from the aspects of power loss and power handling. However, waveguide manufacturing process is difficult and expensive. Recent investigations on substrate integrated waveguides (SIWs), suggest that SIW properties are transitional between microstrip and dielectric-filled waveguide properties. Significant advantages of SIW devices are light weight, low cost, low loss and easy integration with planar circuits [21], [22].

SIW LWAs have been the subject of many studies in recent years [22]–[27] and structures such as those presented in [23] and [26] have achieved the desired results using a simple structure. Various LWAs based on SIW have been proposed, of which the leaky-wave long slot antenna based on SIW is presented in [28]. A straight slot etched on the broadside of a meandering SIW and a meandering slot etched on the broadside of a straight SIW, are shown in [28] which has realized approximate values of $\mathrm{SLL} = -30$ dB and $\mathrm{cross\text{-}polarization} = -30$ dB, through a simple structure design.

Since the slot is designed in a way to have a symmetrical structure, straight slot LWAs based on SIW can improve the cross-polarization level compared to other structures. By tapering the leakage rate along the antenna length, the SLL can be controlled, and it is desirable for the leaky-mode phase constant to remain unchanged along all the antenna length, so that all parts of the antenna aperture radiate at the same angle [1]. In [29]–[31], a design procedure has been presented for creating a two-dimensional graph based on the effects of different parameters of the structure on the phase constant and the leakage rate. This graph can be used for creating a design with different $\alpha$ values and fixed $\beta$ values. In [28], changing the distance of the slot from the sidewall can vary both leakage rate and phase constant and as a result the antenna design cannot realize a variable leakage rate and a fixed phase constant. However based on the results in [31], the antenna could be designed with a fixed phase constant in order to obtain the desired radiation pattern.

The proposed antenna structure is an SIW long slot LWA. By transversely varying the ridge located beside the SIW sidewalls, a variable leakage rate and a fixed phase constant can be obtained to realize the desired SLL. Since the slot is symmetrical, a low cross-polarization is also achieved.

Different theoretical and numerical approaches have been employed in computing the leaky-wave structures properties. Magnetic field integral equation in [25] and [32], [33] and method of moment in [17] and [34], [35] are commonly used to analyze LWAs.

Another method is an analytical procedure based on a transverse equivalent network (TEN) for the LWA. This approach is accurate and not time-consuming. This method is used to analyze slit asymmetric waveguides and versatile LWAs based on stub-loaded rectangular waveguides in [6] and [36], [37]. Also in [29], the method of moment is applied in conjunction with a multimodal TEN.

In this paper, a dispersion relation is obtained from the resonance of the TEN. It has to be noted that the presented TEN for the proposed structure is created based on the obtained results in [36]–[39]. A suitable and accurate TEN is derived and the complex dispersion relation of the LWA is obtained.

The theoretical models and circuit models used to analyze the proposed antenna are outlined in Section III. Section IV presents simulation results of the HFSS software for the uniform long slot LWA and conformity between the leakage rate ($\alpha$) and phase constant ($\beta$) resulted from the TEN models and HFSS simulations is observed.

In Section V, the LWA is designed based on the curves in Section IV in order to realize an optimum SLL and cross-polarization. Section VI shows that measurement results are consistent with the simulation results and theoretical analysis.

## II. Principle of Operation of the LWA Based on Ridged SIW (RSIW)

In order for the antenna to radiate, the slots should cut off the current distribution on the surface of the SIW. In other words, the field distribution on both sides of the slots should be unlike. When the SIW is operating in the dominant mode, the electric current is not cut off, if the slots are located in center of the upper plane of the antenna and the fields on both sides of the slots are symmetrical, thus the slots wouldn't radiate. Whereas distanced slots from the center of the upper plane of the SIW would make radiation possible. Lack of symmetry in the structure would improve the SLL, but on the other hand the field distribution in the slot aperture increases the cross-polarization.

However if the slot is located in the center of the upper plane of the SIW, and the field distribution in the SIW is controlled in a way that provides the needed conditions, radiation would be possible. In fact, instead of varying the slot position, the field distribution inside the SIW is controlled. In this state the slot can radiate while the cross-polarization is proper. By controlling slotted SIW antenna features, the required leakage rate and consequently the proper SLL can be obtained.

To vary the symmetry of the field distribution inside the SIW, a ridge can be placed alongside one of the SIW sidewalls. The electric field at the ridge would be zero, and it would not be symmetrical with respect to the SIW center. $\alpha$ and $\beta$ vary based on the width and height of the ridge. Considering the fabrication process, and avoiding an increase in the SIW layers, the height of the ridge is assumed to be fixed, and the ridge is only transversely varied. Transversal variation of the ridge would change the field distribution inside the SIW and as a result $\alpha$ is varied, but $\beta$ will vary as well and consequently the desired radiation pattern cannot be obtained. In order to eliminate this problem, in the proposed antenna a ridge was used alongside each sidewall of the SIW structure, instead of just one of the sidewalls, as shown in Fig. 1. Again in this situation, having assumed a fixed ridge height, using ridges with different widths can vary the field distribution at the slot. Controlling the ridge width would result in a variable $\alpha$ and a fixed $\beta$, which in turn would make the desired SLL feasible.

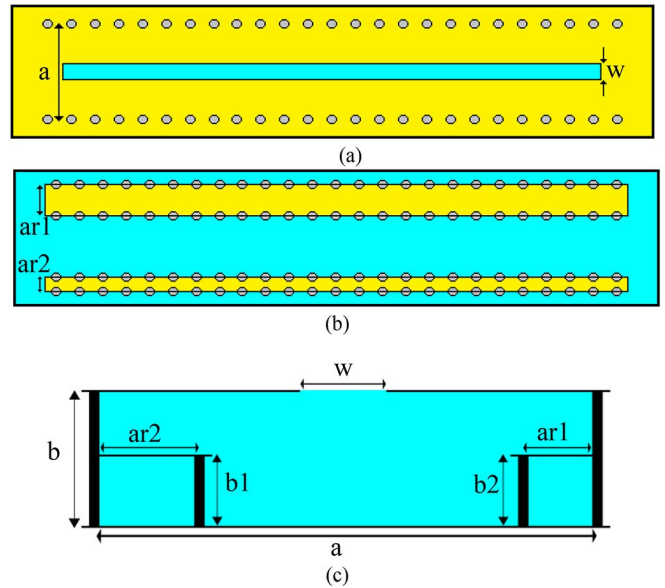

Fig. 1. Geometry of the RSIW LWA (a) top view, (b) middle view, and (c) cross-section view.

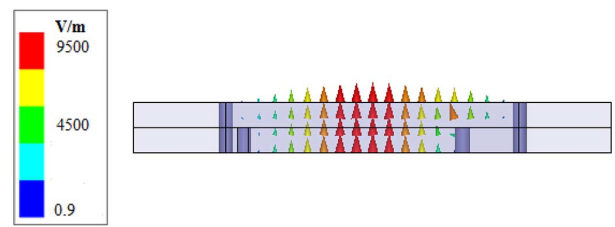

Fig. 2. Electric field distribution in the cross-section of the proposed structure.

To better understand the behavior of the SIW slot antenna, the distribution of the electric field is shown in Fig. 2. It shows the distribution of the electric field in the cross-section of the SIW with asymmetric ridges and without the slot on the upper wall. As can be seen, the electric field has been distributed asymmetrically. As a result the slot can radiate, though it is located in the center of the upper plane of the RSIW.

## III. Dispersion Relation of the Leaky-Mode

The radiation mechanism of an LWA is defined by the complex longitudinal propagation constant of the excited leaky-wave mode in the antenna. The propagation constant is formed by the phase constant $\beta$ and the leakage rate $\alpha$, as [1]

$$k = \beta - j\alpha \quad (1)$$

From the phase and leakage constants, the angle of maximum radiation from the broadside direction ($\theta_\mathrm{m}$) and the 3-dB beamwidth ($\Delta\theta$) can be determined as

$$\sin\theta_m \cong \frac{\beta}{k_0} \quad (2)$$

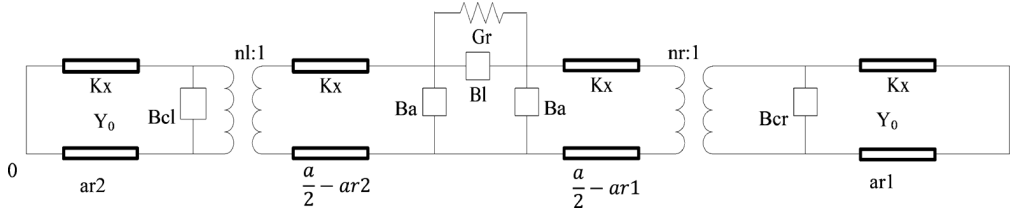

Fig. 3. TEN of the RSIW long slot antenna.

$$\Delta\theta \cong \frac{1}{\frac{L_A}{\lambda_0}\cos\theta_m} \approx \frac{\alpha/k_0}{0.183 \cdot \cos\theta_m} \qquad (3)$$

Where $\lambda_0$ is the free-space wavelength and $L_A$ is the length of the LWA. Both $\alpha$ and $\beta$ are in radians in (2) and (3). The length of the antenna has been chosen so that 90% of the power is radiated, while the remaining 10% is absorbed by a matched load.

$\alpha$ and $\beta$ are known as a function of the antenna geometry and frequency. From (2) and (3), it is understood that, all the antenna properties can be determined easily by the obtained $\alpha$ and $\beta$. Therefore the LWA can be designed to conform to the desired parameters. Analytical expressions and simulation results that provide such information are presented in Section IV.

## IV. TRANSVERSE EQUIVALENT NETWORK

The theoretical method used to obtain the numerical values proposed in this paper involves an analytical procedure based on an accurate simple transverse *equivalent network* for the slotted RSIW structure. The *dispersion relation* can be calculated by the resonance of this simple TEN. In the TEN used here, the higher order mode is not included. Simple TEN is faster than the multimode TEN, however the multimode TEN has a more accurate response. Since the simple TEN has some limitations, these limitations are accounted for in the design of the proposed structure so that the desired results can be achieved. For instance the width of the ridges should not be too large to make the ridge too close to the center of the SIW. These limitations are discussed in [38].

This TEN means that analytical expressions are possible in *simple closed forms*. As a result, the propagation constant itself is in a simple closed form permitting the numerical values to be calculated *quickly*. This method has been developed in [6] and [36]–[39] for analyzing LWA structures.

The top and middle layers of the proposed structure are shown in Fig. 1. It can be seen that the structure consists of a substrate integrated waveguide with double ridges along the sidewalls and a long slot in the center of the upper plane of the SIW.

The TEN for proposed structure is given in Fig. 3. In order to obtain the TEN of the RSIW antenna, the structure can be divided as follows:

1) Left and right sides of the SIW with respect to the center of the slot, where a ridge is located beside a sidewall. As an example, the right side can be considered as two waveguides (transmission lines), with two different heights and widths, where (b-b2)*(ar2) are the dimensions of the first waveguide (transmission line), and (b)*(a-ar2) are the dimensions of the second waveguide [6]. Each one of these waveguides can be assumed as a transmission line, connected by a transformer.

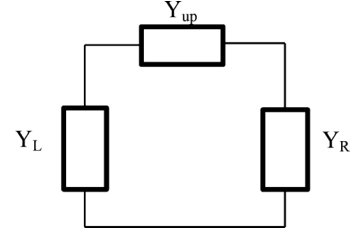

Fig. 4. Reduced network for the proposed TEN.

The transformer coefficient (nr:1) is also obtained based on the impedances before and after the discontinuity [35]. The discontinuity susceptance for the ridge can also be calculated [36]. The left side of the RSIW can be dealt with similarly as the right side. It is the asymmetric structure of the left and right ridges which makes radiation possible.

2) A slot with a width of "W" on the upper plane of the RSIW [37].

Each sidewall of the SIW, which is composed of vias, has been considered as a single ideal continuous plate, and is shown as a short circuit in the equivalent circuit.

The method used for creating the equivalent network is similar to the method used in [6] for an LWA, with the central element in the TEN representation.

The previously proposed TEN can be reduced to the simple circuit shown in Fig. 4.

The transverse resonance equation has been derived from the reduced network, where $Y_L$ and $Y_R$ are the total admittances introduced by the left and right arms. $Y_L$ and $Y_R$ are obtained by

$$Y_{\text{up}} = jB_a + G_r \qquad (4)$$

$$Y_R = Y_{lr}\frac{Y_0 + Y_{lr}\cot\left(k_x(a/2 - ar2)\right)}{Y_{lr} + Y_0\cot\left(k_x(a/2 - ar2)\right)} + jB_L \qquad (5)$$

$$Y_L = Y_{Ll}\frac{Y_0 + Y_{Ll}\cot\left(k_x(a/2 - ar1)\right)}{Y_{Ll} + Y_0\cot\left(k_x\cot(k_x(a/2))\right)} + jB_L \qquad (6)$$

where

$$Y_{lr} = \frac{1}{n_{cr}^2}\left\{jB_{cr} + jY_0\cot\left(k_x(ar2)\right)\right\} \qquad (7)$$

$B_a, n_c$ and $B_L$ are defined in the Appendix and can be replaced in these formulas by their corresponding definitions

$$Y_{Ll} = \frac{1}{n_{cl}^2}\left\{jB_{cl} + jY_0\cot\left(k_x(ar1)\right)\right\}. \qquad (8)$$

The dispersion relation for the propagation behavior is obtained by solving

$$Y_L + Y_R + \frac{Y_L Y_R}{Y_{\text{up}}} = 0. \qquad (9)$$

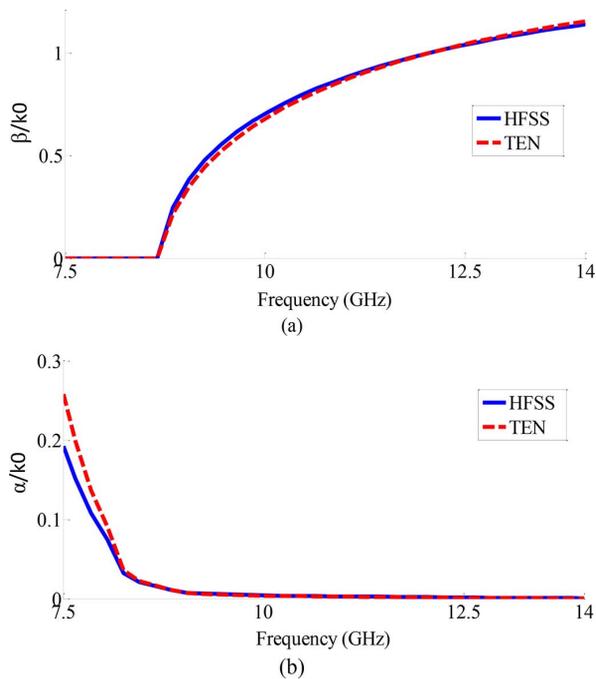

Fig. 5. Simulated complex propagation constant for the perturbed $TE_{10}$ mode.

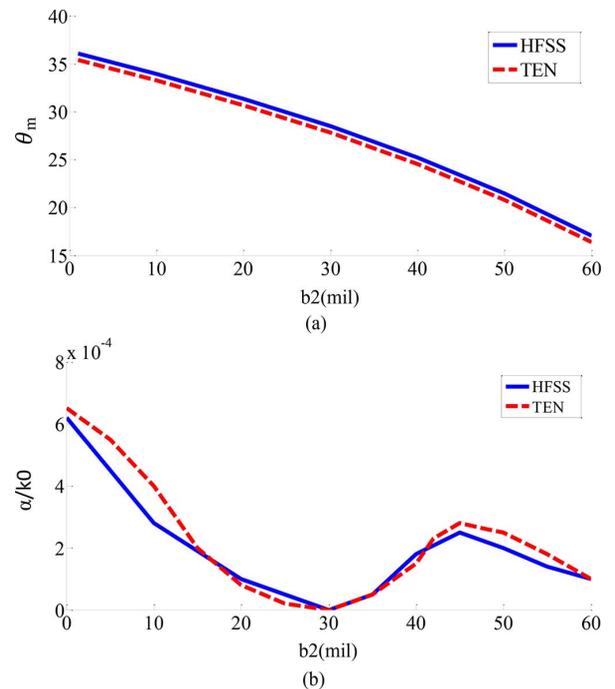

Fig. 6. Leakage rate and beam angle variations versus the height of the ridge at 10 GHz. (b1 = 30 mil).

Equation (9) is the final form of the transverse resonance equation. The numerical solution of (9) easily results in a simple closed form equation for $\alpha$ and $\beta$ for all geometrical parameters.

## V. RESULTS FOR THE PROPAGATION BEHAVIOR OF THE UNIFORM LWA BASED ON RSIW

Numerically solving (9) which is obtained from the TEN, the normalized phase and leakage constants for the geometrical parameters of the proposed antenna can be calculated. In this section, using HFSS simulations, $\alpha$ and $\beta$ for different antenna dimensions and at different frequencies are obtained and compared to the results obtained from the TEN, to figure out the preciseness of this theoretical method. For calculating the antenna radiation properties, [10] is used.

### A. Variations of $\alpha$ and $\beta$ Versus Frequency

LWA is based on the RSIW dominant mode ($TE_{10}$) which is perturbed by removing a slot from its upper plane. Therefore, the dimensions of the RSIW based LWA must be selected for a given frequency of operation.

Fig. 5(a) and (b) shows the variations of $\beta/k_0$ and $\alpha/k_0$ versus frequency, the mentioned figures were created using HFSS simulation and TEN. The results would suggest that HFSS simulation results and that of TEN are well matched, exhibiting the preciseness of the TEN method.

Using (3) results, beam direction from broadside ($\beta/k_0 = 0$) to forward end fire ($\beta/k_0 = 1$) can be obtained. Also using (2), the 3-dB beamwidth can be calculated.

Looking at the curve for $\beta/k_0$, three separate frequency ranges can be seen. At the highest range of frequencies, when $f$ is above 12.5 GHz, the value of $\beta/k_0$ exceeds unity, because the guided wave is a surface wave (slow wave). At that frequency range, it was found that $\alpha/k_0 = 0$. In the second frequency range, from about 8.5 GHz to 12.5 GHz, the guided wave is above cutoff, which is a fast wave. The mode is leaky within that range, and the value of $\alpha/k_0$ decreases as the frequency increases. In the third frequency range which is below 8 GHz, the mode goes below cutoff. The $a/k_0$ value continues to increase strongly, but this region cannot be used for antenna applications, as the wave does not radiate and is reflected back to the source.

The variations in the values of the beam angle and the beamwidth as a function of frequency in the range above cutoff can be computed with (2) and (3). The beam angle changes quickly with frequency, making frequency scan in elevation effective for LWAs. It is possible to cover essentially the whole angular range from broadside to forward end fire by varying the frequency around the central frequency of 10 GHz.

Good agreement is observed between TEN results and HFSS simulations, therefore the TEN approach is validated.

### B. Dependence of $\alpha$ and $\beta$ on Geometrical Parameters

The angle of the maximum beam is obtained using the value of the phase constant and the 3-dB beamwidth related to the value of the leakage rate. Therefore, the behaviors of $\beta$ and $\alpha$ vary as the dimensions of the proposed antenna are changed.

The four best candidates to vary in order to modify $\alpha$ and $\beta$ are the relative slot width, W, the difference in the height of the ridges, the distance between the ridges, and transversal ridge changes, i.e., their distance from the sidewalls, as seen in Fig. 1.

Fig. 6 shows the variations of $\theta_m$ and $a/k_0$ when the height of one of the ridges varies from 0 to b while the height of the other ridge remains constant.

It is seen that varying b2/b1 changes both $\beta/k_0$ and $\alpha/k_0$ simultaneously, increasing one while decreasing the other.

When b1 equals b2, making the structure symmetrical, radiation from the slot stops and $\alpha/k_0 = 0$.

Transversely modifying the ridges, the leakage rate can be controlled within the range of zero to the maximum value. The smooth leakage rate variation from the maximum value, in a highly asymmetric location of the ridges, to zero radiation, with fully symmetrical ridges can easily be seen.

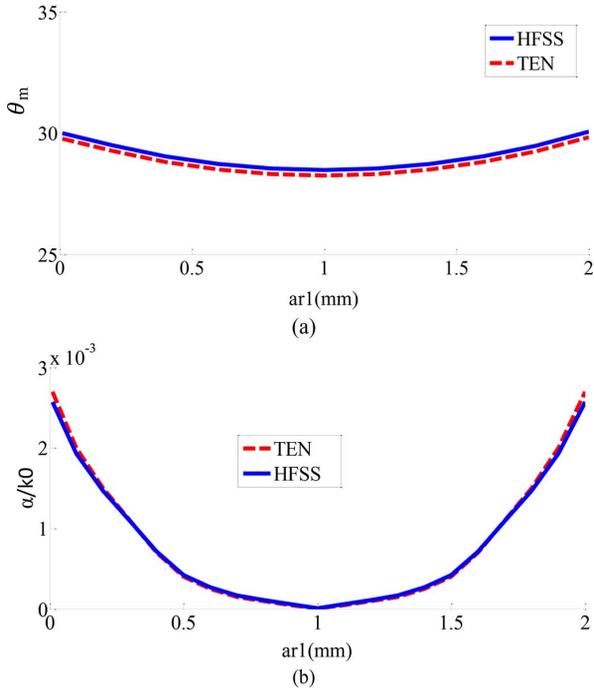

Fig. 7. Leakage rate and angle of beam varying the ar1 with constant offset of the width of the ridges at 10 GHz. (b1 = b2 = 30 mil).

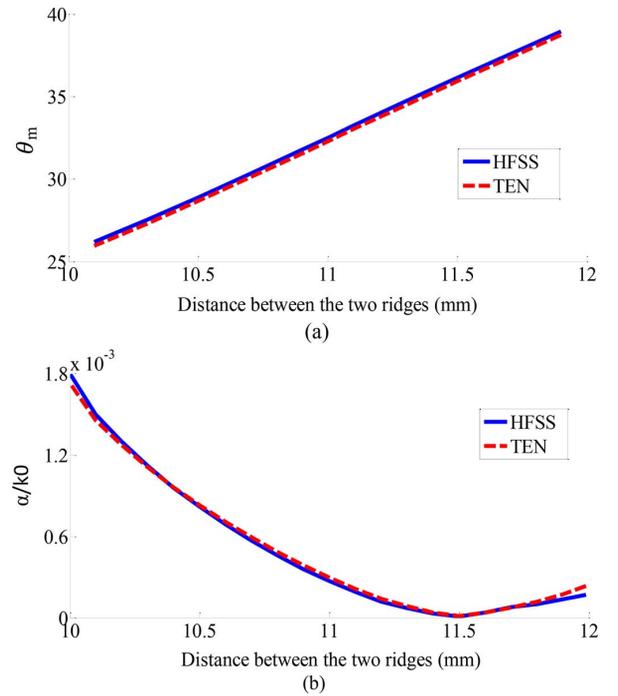

Fig. 8. Leakage rate and beam angle variations versus the distance between two ridges at 10 GHz. (b1 = 30 mil).

In order to better illustrate the asymmetric radiation mechanism, Fig. 7 shows the normalized phase and leakage constants obtained from the perturbed mode simulations, as a function of the transversal ridge changes ar1 and ar2 for a constant ridge height of b1 = b2 = 30 mm. The frequency of analysis is 10 GHz. In this situation, the distance between the two ridges ($\delta$ = a − (ar1 + ar2)) is fixed and as a result ar2 decreases when ar1 is increased. Consequently, in Fig. 7 when ar1 is known, the corresponding ar2 can be found and this is possible due to the fixed $\delta$ and $a\Theta_m$ with little changes.

The behavior of $\alpha$ and $\beta$ as a function of the distance between the ridges ($\delta$) is shown in Fig. 8. The value of $\alpha/k_0$, varies smoothly over a very wide range of values as $\delta$ is changed. The value of $\beta/k_0$ slightly varies over this range, though it was expected to be better.

A 1.5-degree variation in beam angle, over the full range of $\delta$ can easily be seen, whereas the beamwidth varies from zero to a maximum value of about 2.30, which is not so large. The beamwidth can vary in a quite large range while the beam angle remains fully constant.

## VI. DESIGN OF THE TAPERED ANTENNA FOR SLL REDUCTION

In order to control the SLL, the designer must be able to control the leakage rate from zero to a maximum value, and keep the phase constant unchanged along the length of the tapered antenna, so that all parts of the antenna aperture radiate at the same angle. Therefore, the variation of the normalized phase and leakage constants of the proposed antenna must be studied as a function of different geometrical parameters of the designed antenna at the operating frequency at which these parameters can change the value of $\alpha$ while maintaining $\beta$ unchanged.

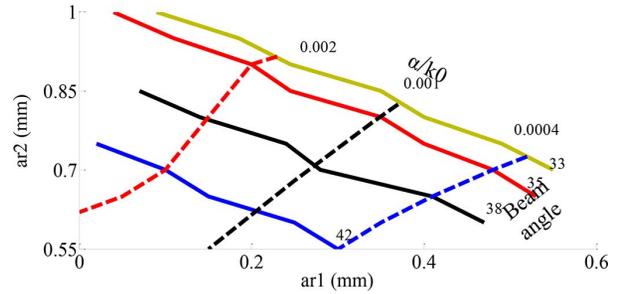

Fig. 9. Leaky-mode contour curve.

It was shown in the Section V, that varying b1 and b2, $\delta$ and ar1 and ar2, would vary both $\alpha$ and $\beta$, thus, none of them can be used alone to access a desirable SLL. On the other hand, if the ridge height variation is to be used for tapering, a multilayer structure should be used, which makes the structure voluminous and the layers are discontinued as the height varies, therefore assuming b1 = b2, $\delta$ and ar1 and ar2 changes would be considered in the design procedure.

For this purpose, In [29]–[31], the procedure for creating a graph is presented which can be used for determining the proper dimensions of the designed structure in order to obtain different $\alpha$ values and fixed $\beta$ values. Based on this graph the leaky-mode contour curve plots shown in Fig. 9 are used. These curves were obtained from HFSS simulations at 10 GHz, by computing the leaky-mode complex propagation constant for w = 2 mm and b1 = b2 = 30 mm, and varying ar1 and ar2. The continuous lines define the pointing angles and the dashed lines define constant leakage rates.

The illumination amplitude function along the antenna aperture has to follow a Taylor distribution with a −35-dB SLL. The required aperture distribution and the calculated radiation pattern are reported in [28]. The variation of the leaky-mode leakage rate along the antenna length ($\alpha(z)$) can be computed from the illumination function ($A(l)$), according to (10) [1]

$$\alpha(z) = \frac{\frac{1}{2}|A(l)|^2}{\frac{1}{\eta}\int_0^L |A(z)|^2\, dz - \int_0^l |A(z)|^2\, dz} \quad (10)$$

where $\eta$ is the antenna efficiency which is determined by the ratio between the amount of the power radiated along the antenna length and the total power injected to the structure. Table I, provides the values of ar1 and ar2 at different positions of an antenna with a length of 13 wavelengths, for achieving the optimum aperture distribution, based on (10). In this case, at any point of the aperture the radiation is in a fixed direction, and the desired leakage rate helps to realize a proper SLL.

TABLE I
THE VALUES OF AR1 AND AR2 FOR $SLL = -35$ dB AND $\theta_m = 40°$

| z(mm) | ar1(mm) | ar2(mm) |
|---|---|---|
| 0 | 0.55 | 0.625 |
| 13λ/20 | 0.6 | 0.65 |
| 2*13λ/20 | 0.59 | 0.66 |
| 3*13λ/20 | 0.57 | 0.68 |
| 4*13λ/20 | 0.55 | 0.7 |
| 5*13λ/20 | 0.53 | 0.73 |
| 6*13λ/20 | 0.455 | 0.77 |
| 7*13λ/20 | 0.4 | 0.8 |
| 8*13λ/20 | 0.35 | 0.85 |
| 9*13λ/20 | 0.27 | 0.885 |
| 10*13λ/20 | 0.22 | 0.915 |
| 11*13λ/20 | 0.185 | 0.95 |
| 12*13λ/20 | 0.1425 | 0.975 |
| 13*13λ/20 | 0.1 | 0.995 |
| 14*13λ/20 | 0.09 | 1 |
| 15*13λ/20 | 0.15 | 0.97 |
| 16*13λ/20 | 0.21 | 0.94 |
| 17*13λ/20 | 0.28 | 0.88 |
| 18*13λ/20 | 0.42 | 0.79 |
| 19*13λ/20 | 0.495 | 0.75 |
| 13λ | 0.533 | 0.727 |

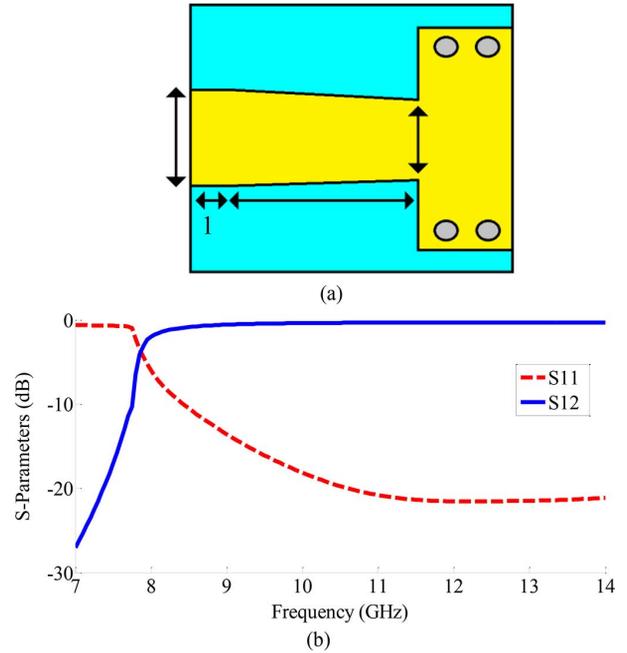

Fig. 10. Input matching structure of the proposed LWA (a) Geometry of the proposed structure (unit: mm) and (b) S-parameters of the proposed structure.

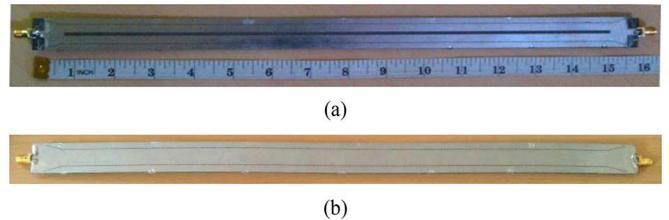

Fig. 11. Fabricated antenna (a) top view and (b) bottom view.

## VII. MEASUREMENT RESULTS

The proposed RSIW based long slot LWA was fabricated on a Rogers5880 substrate with $\varepsilon_r = 2.2$ and $b = 0.8$ mm, for each substrate, as shown in Fig. 11.

Feeding technique is an important part of the design that influences the input impedance and characteristics of the antenna. The microstrip feed is used in this structure to widen the bandwidth of the transmission line to match the needs of the proposed LWA. Simultaneous field matching and impedance matching are important for a transition design. In this structure, for impedance matching between the microstrip line and the SIW, the microstrip line has been tapered from the input impedance of a coaxial cable with an impedance of 50 ohms to the input impedance of the SIW. According to HFSS 70 ohms. Also for field matching between the microstrip line and the SIW, the microstrip line is located on the same plane as that of the top of the SIW. The input and output matching structures are the same as shown in Fig. 10(a). The s-parameters of the input matching are shown in Fig. 10(b).

The measured and simulated normalized copolarization and cross-polarization radiation patterns at the frequencies of 9, 10, and 11 GHz are plotted and compared in Fig. 12(a)–(c), respectively. A good agreement is achieved between the measured and the simulated radiation patterns. It can be seen that SLL reduction is maintained for different scanning angles. Table II, contains values of gain and radiation efficiency of the proposed antenna in frequencies of 9, 10, and 11 GHz.

The scattering parameter of the designed and tapered LWA is plotted in Fig. 13. The measured $S_{11}$ parameter is in a good agreement with HFSS simulations. The measured reflections of the proposed antenna are almost below $-10$ dB in the frequency range of 8.5 to 12.5 GHz.

## VIII. CONCLUSION

The new SIW long slot LWA is studied and realized with controllable radiation properties. To decrease the cross-polarization, the long slot was placed on the upper plane of the SIW. The advantage of the proposed LWA is that the independent ar1 and ar2 parameters provide an unchanged $\beta$ and a controllable $\alpha$. Therefore, the proposed antenna can realize an optimum side lobe level of less than $-28$ dB. A simple dispersion relation for the proposed antenna which is solved numerically to yield its complex dispersion behavior can be derived by the transverse resonance network and the TEN method presented by Pariello and Olinear has been developed for the proposed structure. Also the wavenumbers of these waves are calculated by HFSS simulations. In order to validate the TEN and HFSS simulation results, the proposed antenna operating at 10 GHz has been built and measured, showing a very good agreement between theory and experiments.

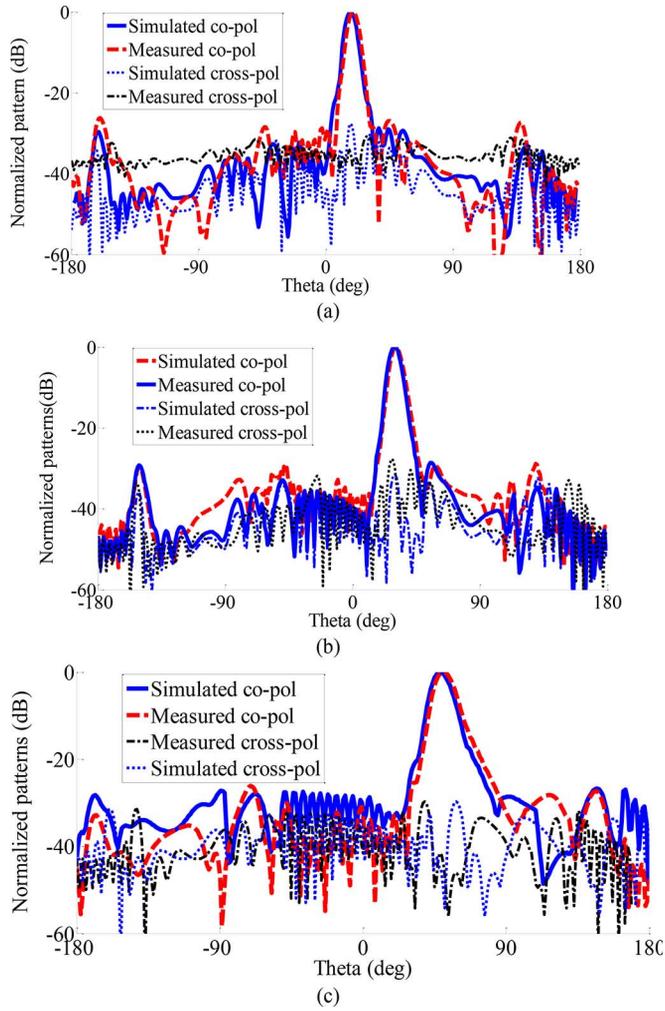

Fig. 12. Normalized radiation patterns for (a) 9 GHz, (b) 10 GHz, and (c) 11 GHz.

TABLE II
GAIN AND EFFICIENCY OF THE PROPOSED ANTENNA

| Frequency (GHz) | Gain(dB) | Efficiency (%) |
|---|---|---|
| 9 | 13.9 | 89.8 |
| 10 | 14.6 | 90.5 |
| 11 | 15 | 91 |

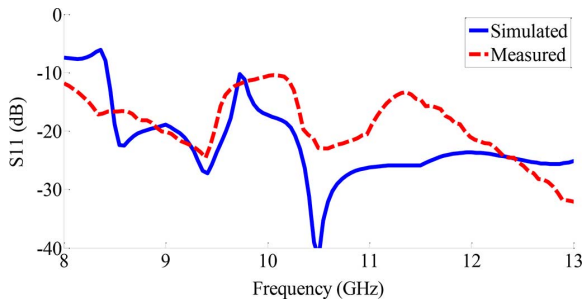

Fig. 13. Measured and simulated of S11.

## APPENDIX

For the equivalent network shown in Fig. 3 and also for the (4)–(7) the different values for the admittances and transformer coefficients are as

$$\frac{B_a}{Y_0} = -2\frac{\pi^2}{16}\frac{b}{\lambda_g}\left(\frac{w_s}{b}\right)^2 J_0^2\left(\frac{\pi w_s}{\lambda_g}\right)$$

$$\frac{B_L}{Y_0} = \frac{1}{nc^2}\left(\frac{2b}{\lambda_g}\right)\left[\ln\left(1.43\frac{b}{w_s}\right) + \frac{1}{2}\left(\frac{2b}{\lambda_g}\right)^2\right]$$
$$+ \frac{\pi^2}{32}\frac{b}{\lambda_g}\left(\frac{w_s}{b}\right)^2 J^2\left(\frac{\pi w_s}{\lambda_g}\right)$$

where $Y_0$ is the characteristic admittance of the SIW and

$$n_c = \frac{\sin(\pi w_s/\lambda_g)}{\pi w_s/\lambda_g}$$

$$n_r = n_l = \sqrt{\frac{b_1}{b}}.$$

These equations are extracted from [6]. Also [39] is used for obtaining $B_{cr}$ and $B_{cl}$.

$$B_{cl} = B_{cr} = T_1 + T_2$$

$$T_1 = \frac{4b_{l/r}}{\lambda_c}\ln\left\{\frac{1-\delta^2}{4\delta}\left(\frac{1+\delta}{1-\delta}\right)^{\{\delta+(1/\delta)\}/2}\right\}$$

$$T_2 = \frac{4b_{l/r}}{\lambda_c}\left\{2\frac{A+A^1+2C}{AA^1-C^2} + \left(\frac{2b_{l/r}}{4\lambda_c}\right)^2\right.$$
$$\left.\times\left(\frac{1-\delta}{1+\delta}\right)^{4\delta}\left(\frac{5\delta^2-1}{1-\delta^2} + \frac{4\delta^2 C}{3A}\right)^2\right\}$$

$$\lambda_c = \frac{2\pi}{k_x}$$

$$\delta = \frac{d}{b_{l/r}}$$

$$A = \left[\frac{1+\alpha}{1-\alpha}\right]^{2\alpha}\left[\frac{1+\sqrt{1-(2b/\lambda_c)^2}}{1-\sqrt{(2b/\lambda_c)^2}}\right] - \frac{1+3\alpha^2}{1-\alpha^2}$$

$$A^1 = \left[\frac{1+\alpha}{1-\alpha}\right]^{2/\alpha}\left[\frac{1+\sqrt{1+(2b1/\lambda_c)^2}}{1-\sqrt{(1-(2b1/\lambda_c)^2}}\right] + \frac{3+\alpha^2}{1-\alpha^2}$$

$$C = \left[\frac{4\alpha}{1-\alpha^2}\right]^2$$

$$\alpha = \frac{b_1}{b}.$$

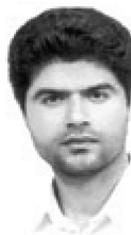

**Alireza Mallahzadeh** (M'12) received the B.S. degree in electrical engineering from Isfahan University of Technology, Isfahan, Iran, in 1999 and the M.S. and the Ph.D. degrees in electrical engineering from Iran University of Science and Technology, in 2001 and 2006, respectively.

He is a member of academic staff, Faculty of Engineering, Shahed University, Tehran, Iran. He has participated in many projects relative to antenna design, which resulted in fabricating different types of antennas for various companies. Also, he is interested in numerical modeling, and microwaves.

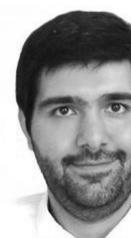

**Sajad Mohammad-Ali-Nezhad** received the B.Sc. degree in electronic engineering from Shahid Chamran University, Ahwaz, Iran, in 2008, and the M.Sc. degree in communication engineering from Shahed University, Tehran, Iran, in 2010, where he is currently working towards the Ph.D. degree in communication engineering.

His main areas of interest are leaky wave antenna, printed circuit antennas, array antennas, phased array antennas, MIMO antennas, RFID tag antenna, frequency selective surface, electromagnetic compatibility, microwave filters and hybrids, and electromagnetic theory.